\documentstyle[12pt]{article}
\begin{document}
\begin{center}
{\bf Bethe-ansatz equations for quantum Heisenberg chains\\
 with elliptic exchange}\\
\end{center}

\begin{center}
{\sc V.I. Inozemtsev}\\
{\it BLTP JINR, 141980 Dubna, Moscow Region, Russia}    \\
\vspace*{1cm}
Abstract\\
\end{center}

\noindent The eigenvectors of the Hamiltonian ${\cal H}_{N}$ of $N$-sites
quantum spin chains with elliptic exchange are connected with the double Bloch
 meromorphic solutions of the quantum continuous elliptic Calogero-Moser
 problem. This fact allows one to find the eigenvectors via the
solutions to the system of highly transcendental equations of Bethe-ansatz
type which is presented in explicit form.\\   

It is known [1] that for a one-parameter set ${\cal H}_{N}$ of linear
combinations of $N(N-1)/2$ elementary transpositions $\{P_{jk}\}$,
${\cal H}_{N}={J\over 2}\sum_{1\leq j\neq k}^{N} \wp(j-k)P_{jk}$
 at arbitrary natural $N\geq 3$, one can construct
a variety $\{I_{m}\}$ $(3\leq m\leq N)$ of operators which commute
with ${\cal H}_{N}$. Being applied to $SU(2)$ spin representations of
the permutation group, this proves the integrability of 1D periodic
spin chains with elliptic short-range interaction and the Hamiltonian
$${\cal H}^{(s)}={J\over4}\sum_{1\leq{j}\neq{k}\leq{N}}h(j-k)
(\vec\sigma_{j}\vec\sigma_{k}-1),
\eqno(1)$$
where 
 $$h(j)=\left({\omega\over\pi}\sin{\pi\over\omega}\right)^2
\left[\wp_{N}(j)+{2\over\omega}\zeta_{N}\left({\omega\over2}\right)\right],
\eqno(2)$$
where $\wp_{N}(x)$, $\zeta_{N}(x)$ are the Weierstrass functions defined
on the torus $T_{N}={\bf C}/{\bf Z}N+{\bf Z}\omega$, $\omega=i\alpha$,
$\alpha\in{\bf R}_{+}$ is a free parameter.

 The symmetry of two limiting cases of this one-parameter model, i.e.
 the Bethe lattice
with nearest-neighbor interaction [2] ($\alpha\to 0$) and long-range
$\left({N\over\pi}\sin{{\pi j}\over{N}}\right)^{-2}$
exchange [3] ($\alpha\to\infty$), is now well understood and regular
procedures of finding eigenvectors are described in the literature [4-7]. 
At present, a number of impressive results are known for both these models.
In particular, they include the additivity of the spectrum under proper
choice of "rapidity" variables [2,3], the
 description of underlying symmetry [4,5], construction of thermodynamics
in the limit $N\to\infty$ [9,10].
However, all that still cannot be applied to
the general elliptic case.

In the paper [8], I have shown that there is a remarkable connection
between the eigenvectors of the Hamiltonian of the above model with $M$ down
spins and double Bloch meromorphic solutions to the quantum continuous
elliptic Calogero-Moser problem at the special value of the coupling
constant, i.e. the eigenfunctions of the {\it differential} operator
$$H= -{1\over 2}\sum_{j=1}^{M}{{\partial^2}\over{\partial x_{j}^2}}+
\sum_{j\neq k}^{M}\wp_{N}(x_{j}-x_{k}). \eqno(3)$$
This allows one in principle to find an ansatz for the eigenvectors and
even try to describe them completely if the solutions to (3) are known.
This has been done in the simplest nontrivial case $M=3$ in [11], where
I have used the result for three-particle elliptic Calogero-Moser problem
[12].

 At that time, the explicit form of the eigenfunctions of (3) at $M>3$
has not been known. The situation has been changed after publishing of seminal
paper [13] where these eigenfunctions have been obtained in the process
of constructing solutions to the elliptic Knizhnik-Zamolodchikov-Bernard
equations. It has been a main motivation for this paper in which I shall
describe the complete set of the Bethe-ansatz-type equations for the
eigenvectors of (1) at {\it arbitrary} $M\leq N/2$.

The Hamiltonian (1) commutes with the operator of total spin $\vec
{\bf S}={1\over2}
\sum_{j=1}^{N}\vec \sigma_{j}$. Then the eigenproblem for it is
decomposed into the problems in the subspaces formed by the common
 eigenvectors of ${\bf S}_{3}$ and $\vec {\bf S}^2$ such that $S=S_{3}=
N/2-M$, $0\leq{M}\leq[N/2]$,
$${\cal H}^{(s)}\vert\psi^{(M)}>=E_{M}\vert\psi^{(M)}>.
\eqno(4)$$
The eigenvectors $\vert \psi^{(M)}>$ are written in the usual form
$$\vert\psi^{(M)}>=\sum_{n_{1}..n_{M}}^{N}\psi_{M}(n_{1}..n_{M})
\prod_{\beta=1}^{M}
s_{n_{\beta}}^{-}\vert0>,
\eqno(5)$$
where $\vert0>=\vert\uparrow\uparrow...\uparrow>$ is the ferromagnetic ground state with all spins up and the summation is taken over all combinations of integers $\{n\}\leq N$ such that
$\prod_{\mu<\nu}^{M}(n_{\mu}-n_{\nu})\neq0$. The substitution of (5) into
(4) results in the {\it lattice} Schr\"odinger equation for completely
symmetric wave function $\psi_{M}$
$$\sum_{s\neq n_{1},..n_{M}}^{N}\sum_{\beta=1}^{M}\wp_{N}(n_{\beta}-s)
\psi_{M}(n_{1},..n_{\beta-1},s,n_{\beta+1},..n_{M})$$
$$+
\left[\sum_{\beta\neq\gamma}^{M}\wp_{N}(n_{\beta}-n_{\gamma})-{\cal E}_{M}
\right]\psi_{M}(n_{1},..n_{M})=0.
\eqno(6)$$
The eigenvalues $\{E_{M}\}$ are given by
$$E_{M}=J\left({\omega\over\pi}\sin{\pi\over\omega}\right)^{2}
\left\{{\cal E}_{M}+{2\over\omega}\left[{{2M(2M-1)-N}\over4}\zeta_{N}\left(
{\omega\over2}\right)-M\zeta_{1}\left({\omega\over2}\right)\right]
\right\},
\eqno(7)$$
where $\zeta_{1}(x)$ is the Weierstrass zeta function defined on the torus
$T_{1}={\bf C}/{\bf Z}+ {\bf Z}\omega$.

The solutions to (6) can be found with the use of the following ansatz
for $\psi_{M}$:
$$
\psi_{M}(n_{1},..n_{M})=\sum_{P\in\pi_{M}}\varphi_{M}^{(p)}(n_{P1},..n_{PM}),
\eqno(8)$$
$$\varphi_{M}^{(p)}(n_{1},..n_{M})=\exp\left(-i\sum_{\nu=1}^{M}\tilde p_{\nu}
n_{\nu}
\right)
\chi_{M}^{(p)}(n_{1},..n_{M}),
\eqno(9)$$
where
$$\tilde p_{\nu}=p_{\nu}-2\pi N^{-1}l_{\nu}, \qquad l_{\nu}\in {\bf Z}
, \eqno(10)$$
 $\pi_{M}$ is the group of all permutations $\{P\}$ of the numbers from 1 to $M$
and $\chi_{M}^{(p)}$ is some special solution to the {\it continuum}
quantum many-particle problem
$$\left[-{1\over2}\sum_{\beta=1}^{M}{{\partial^2}\over{\partial x_{\beta}^2}}
+\sum_{\beta\neq\lambda}^{M}\wp_{N}(x_{\beta}-x_{\lambda})-{\sf E}_{M}(p)\right]
\chi_{M}^{(p)}(x_{1},..x_{M})=0.
\eqno(11)$$
It is specified up to a normalization factor by the particle pseudomomenta
$(p_{1},..p_{M})$. The standard argumentation of the Floquet-Bloch theory shows
that due to perodicity of the potential term in (49) $\chi_{M}^{(p)}$ obeys
the quasiperiodicity conditions 
$$\chi_{M}^{(p)}(x_{1},..x_{\beta}+N,..x_{M})=\exp(ip_{\beta}N)
\chi_{M}^{(p)}(x_{1},..x_{M}),
\eqno(12)$$
$$\chi_{M}^{(p)}(x_{1},..x_{\beta}+\omega,..x_{M})=\exp
(2\pi iq_{\beta}(p)+ip_{\beta}\omega)
\chi_{M}^{(p)}(x_{1},..x_{M}), \quad 0\leq{\Re}e(q_{\beta})<1,
\eqno(13)$$
$$
1\leq\beta\leq M.
$$
The eigenvalue ${\sf E}_{M}(p)$ is some symmetric function of $(p_{1},..p_{M})$.
As will be seen later, the set $\{q_{\beta}(p)\}$ is also completely
determined by $\{p\}$.

 It turns out [8] that the equation (6) with the use of (8),(9)
 can be recast in the form
$$
\sum_{P\in\pi_{M}}\left[-{1\over2}\sum_{\beta=1}^{M}
\left({\partial\over{\partial n_{P\beta}}}-f_{\beta}(p)\right)^2+
\sum_{\beta\neq\gamma}^{M}\wp_{N}(n_{P\beta}-n_{P\gamma})-
\right.$$
$$\left.{\cal E}_{M}+
\sum_{\beta=1}^{M}\varepsilon_{\beta}(p)\right]\varphi^{(p)}(n_{P1},..
n_{PM})=0,
\eqno(14)$$
where
$$f_{\beta}(p)=2\tilde q_{\beta}(p)\zeta_{1}(1/2)
-\zeta_{1}(\tilde q_{\beta}(p)),
\eqno(15)
$$
$$\varepsilon_{\beta}(p)={1\over2}\wp_{1}(\tilde q_{\beta}(p)),
\eqno(16)$$
$$\tilde q_{\beta}(p)= q_{\beta}(p)+{{l_{\beta}}\over N}\omega.
\eqno(17)$$
where $\wp_{1}(x),\zeta_{1}(x)$ are the Weierstrass functions defined
on the torus $T_{1}={\bf C}/{\bf Z}+ {\bf Z}\omega$.

Turning to the definition (9) of $\varphi^{(p)}$,
one observes that each term of the left-hand side of (14) has the same structure
as the left-hand side of the many-particle Schr\"odinger equation (11) and
vanishes if ${\cal E}_{M}$ and $f_{\beta}(p)$ are chosen as
$$f_{\beta}(p)=-i\tilde p_{\beta},\qquad \beta=1,..M,
\eqno(18)$$
$${\cal E}_{M}={\sf E}_{M}(p)+\sum_{\beta=1}^{M}\varepsilon_{\beta}(p).
\eqno(19)$$

One can see from (15-19) that it remains now to find the explicit dependence
of $\{q\}$ and ${\sf E}_{M}$ on $\{p\}$. It can be done by using the results
given in [13] where the explicit form of $\chi_{M}^{(p)}(x)$ has been
indicated. In suitable notations, it reads
$$\chi_{M}^{(p)}(x)\sim\exp(i\sum_{\beta=1}^{M}p_{\beta}x_{\beta})
\sum_{s\in \pi_{m}}l(s)\prod_{j=1}^{m}\tilde\sigma_{\sum_{k=1}^j (x_{c(s(k))
}-x_{c(s(k))+1})}(t_{s(j)}-t_{s(j+1)}),
\eqno(20)$$
where $m=M(M-1)/2$, $c$ is non-decreasing function $c:\{1,..,m\}\to
\{1,..,M-1\}$ such that $\vert c^{-1}\{j\}\vert=M-j$,
  $l(s)$ is an integer which is defined for the
permutation $s$ by the relation $x_{c(s(1))+1}\partial/\partial x_{c(s(1))}
...x_{c(s(m))+1}\partial/\partial x_{c(s(m))}x_{1}^{M}=$ $l(s)(x_{1}...x_{M})
$, $\{t\}$ is a set of $m$ complex parameters obeying $m$ relations [13]
$$\sum_{l:\vert c(l)-c(j)\vert =1}\rho(t_{j}-t_{l})-2\sum_{l:l\neq j,c(l)=
c(j)}\rho(t_{j}-t_{l})+M\delta_{c{j},1}\rho(t_{j})=i(p_{c(j)}-p_{c(j)+1}),
\eqno(21)$$
$$\rho(t)=\zeta_{N}(t)-{2\over N}\zeta_{N}(N/2)t,$$
and 
$$\tilde\sigma_{w}(t)=\exp((2/N)\zeta_{N}(N/2)wt){{\sigma_{N}(w-t)}\over
{\sigma_{N}(w)\sigma_{N}(t)}},$$
$\sigma_{N}$ being the Weierstrass sigma function on ${\bf T}_{N}$.
The elementary building blocks of the $\chi$ function obey the useful
quasiperiodicity relations
$$
\tilde\sigma_{w+N}(t)=\tilde\sigma_{w}(t),\qquad
\tilde\sigma_{w+\omega}(t)=e^{2\pi it/N}\tilde\sigma_{w}(t).
\eqno(22)$$
  One can see that in this construction the color function $c(j)$ is of
crucial role. It is useful to write it explicitly. Namely, define for
every $k$=1,..$M-1$ the segment $S_{k}$
$$ {{(k-1)(2M-k)}\over 2}+1\leq j\leq{{k(2M-k-1)}\over 2}. \eqno(23)$$
Then some calculation shows that
$$c(j)=k \qquad {\rm if}\quad j\in S_{k}
.\eqno(24)$$
  
The main advantage of the explicit form of $\chi$ function is that it
allows to find the second set of relations between the Bloch factors
$\{p\}, \{q\}$. It is easy to see from (21) that $\{p\}'s$ in the definitions
(12) and (20) are the same. The problem consists in calculation of $\{q\}$.
To do this, it is not necessary to analyze each term in the sum over 
permutations in (20) since all of them must have the same Bloch factors.
It is convinient to choose the term which corresponds to the permutation 
$$s_{0}: \quad s_{0}(j)=m+1-j,\quad j=1,..m.$$
After some algebra, one finds that this permutation gives nontrivial
contribution to the sum (20) with $l(s_{0})=M!(M-1)!...2!$. Moreover,
with the use of explicit form of the color function (23-24) one finds
$$c(s_{0}(l))=M-q\quad {\rm if}\quad q(q-1)/2+1\leq l\leq q(q+1)/2.$$
 Now the problem of calculation of the second Bloch factors reduces, due
to second relation (21), to some long and tedious, but in fact simple calculations
of the product of factors which various $\tilde\sigma$ functions acquire
under changing arguments of $\chi$ function to the quasiperiod $\omega$.
The final result is surprisingly simple,
$$q_{\beta}(p)=N^{-1}\left(\sum_{l:c(l)=\beta}t_{l}-\sum_{l:c(l)=\beta -1}
t_{l}\right),\qquad 1<\beta<M-1,\eqno(25)$$
with the first and second term being omitted for $\beta=M$ and $\beta=1$.
 
The equations (25), together with (18) and (21), form a closed set
for finding Bloch factors $\{p\},\{q\}$ at given integers
$\{l_{\beta}\}\in {\bf Z}/M{\bf Z}$ and determining the eigenvalues of
the spin Hamiltonian (1,2) completely.
The corresponding eigenvalue of the continuum $M$-particle operator (11) is
given by [13]
$${\sf E}_{M}(p)={{2M(M-1)}\over N}\zeta\left({N\over 2}\right)
+\sum_{\beta=1}^{M}p_{\beta}^{2}/2$$ 
$$-{1\over 2}\left[\sum_{k<l}^{m}(2\delta_{c(k),c(l)}F(t_{k}-t_{l})-
\delta_{\vert c(k)-c(l)\vert,1}F(t_{k}-t_{l}))-M\sum_{c(k)=1}F(t_{k})
\right],\eqno(26)$$
where
$$
F(t)=-\wp_{N}(t)+(\zeta_{N}(t)-2/N\zeta_{N}(N/2))^2 +4/N\zeta_{N}(N/2).
$$
This allows one to find, via (7) and (19), the explicit form of the
eigenvalues of spin Hamiltonian (1,2). It is worth noting that for their
real calculation one has to solve the Bethe-type equations (18), (21),
(25) at first.

In conclusion, it is demonstrated that the procedure of the exact
 diagonalization of the lattice Hamiltonian with the non-nearest-neighbor
elliptic exchange can be reduced in each sector of the Hilbert space with
given magnetization to the construction of the special double quasiperiodic
eigenfunctions of the many-particle Calogero-Moser problem on a
continuous line. The equations of the Bethe-ansatz form appear very
naturally as a set of restrictions to the particle pseudomomenta. The proof
of this correspodence between lattice and continuum integrable models is
based only on analytic properties of the eigenfunctions. One can expect
that the set of spin lattice states constructed by this way is complete.
This is supported by exact analytic proof in the two-magnon case.

The analysis of explicit form of the equations (21) available for
$M=2,3$ shows that the spectrum of the lattice Hamiltonian with the
exchange (1) is {\it not}
additive being given in terms of pseudomomenta $\{p\}$ or phases which
parametrize the sets $\{p,q\}$ [11]. For arbitrary $M$, this can be seen
directly from (26). The problem of finding appropriate
set of parameters which gives the "separation" of the spectrum remains
open. It would be also of interest to consider various limits $(N\to
\infty, \alpha\to 0,\infty)$ so as to recover the results of the papers
[2,3] and prove the validity of the approximate methods of asymptotic
Bethe ansatz.        \\
\newpage
{\bf References}
\begin{enumerate}
\item
V.I. Inozemtsev. Lett. Math. Phys. 36,55 (1996)
\item
H. Bethe. Z.Phys. 71,205 (1932)
\item
F.D.M. Haldane. Phys. Rev. Lett. 60,635 (1988); B.S. Shastry, {\it ibid}
60,639 (1988)
\item
L.D. Faddeev, in {\it Integrable models of 1+1 dimensional quantum field
theory}, Elsevier, Amsterdam, 1984
\item
D. Bernard, M. Gaudin, F.D.M. Haldane and V. Pasquier. J.Phys. A26,5219 (1993)
\item
M. Fowler and J. Minahan. Phys. Rev. Lett. 70,2325 (1993)
\item
B.S. Shastry and B. Sutherland. Phys. Rev. Lett. 71,5 (1993)
\item
V.I. Inozemtsev. J.Phys. A28,L439 (1995)
\item
F.D.M. Haldane. Phys. Rev. Lett. 66,1529 (1991)
\item
M. Takahashi. Progr. Theor. Phys. 46,401 (1971)
\item
V.I. Inozemtsev. J.Math. Phys. 37,147 (1996)
\item
J. Dittrich and V.I. Inozemtsev. J. Phys.A: Math. Gen. 26,L753 (1993)
\item
G. Felder and A. Varchenko. Int. Mat. Res. Notices 5,222 (1995)
\end{enumerate}

\end{document}